\documentclass{amsart}

\usepackage[dvips]{graphicx}
\usepackage{amsthm}
\usepackage{amssymb}

\newcommand*{\cn}{\mathop{\mathrm{cn}}\displaylimits}
\newcommand*{\sn}{\mathop{\mathrm{sn}}\displaylimits}
\newcommand*{\dn}{\mathop{\mathrm{dn}}\displaylimits}
\newcommand*{\arccosh}{\mathop{\mathrm{arccosh}}\displaylimits}

\newtheorem{theorem}{Lemma}
\newtheorem{corollary}[theorem]{Corollary}

\begin{document}

\title[Propagation of binary signals]{On the propagation of binary signals in damped mechanical systems of oscillators}

\author{J. E. Mac\'{\i}as-D\'{\i}az}
\address{Departamento de Matem\'{a}ticas y F\'{\i}sica, Universidad Aut\'{o}noma de Aguascalientes, Aguascalientes, Ags. 20100, Mexico}
\address{Department of Physics, University of New Orleans, New Orleans,  LA 70148}
\email{jemacias@correo.uaa.mx}

\author{A. Puri}
\address{Department of Physics, University of New Orleans, New Orleans,  LA 70148}
\email{apuri@uno.edu}


\begin{abstract}
In the present work, we explore efficient ways to transmit binary information in discrete, semi-infinite chains of coupled oscillators using the process of nonlinear supratransmission. A previous work showed that such transmission is possible and, indeed, reliable under the idealistic condition when weak or no damping is present. In this paper, we study a more realistic case and propose the design of mechanical devices in order to avoid the loss of information, consisting on the linear concatenation of several such mechanical systems. Our results demonstrate that the loss of information can be minimized or avoided using such physical structures.
\end{abstract}

\maketitle

\section{Introduction}

The recently discovered process of energy transmission in the forbidden band gap of nonlinear chains \cite{Caputo-Leon-Spire} has opened a wide variety of new physical applications. This phenomenon consists in the sudden increase in the amplitude of wave signals that propagate in a semi-infinite, discrete, nonlinear chain driven at its end by a harmonic disturbance irradiating at a frequency in the forbidden band gap, and it has been studied in sine-Gordon and Klein-Gordon systems \cite{Geniet-Leon}, double sine-Gordon systems \cite{Geniet-Leon2}, Fermi-Pasta-Ulam systems \cite{Khomeriki}, Bragg media in the nonlinear Kerr regime \cite{Leon-Spire}, and even in continuous, nonlinear, bounded media described by undamped sine-Gordon equations driven at one end \cite{Khomeriki-Leon}.

From a mathematical point of view, the theory that describes the mechanism sustaining the phenomenon of nonlinear supratransmission has not been entirely unveiled in general. In fact, most of the analytical results on undamped, discrete systems that we posses nowadays rely on the use of approximation techniques such as the rotating wave approximation \cite{Khomeriki}, or the study of the continuous limiting case \cite{Geniet-Leon}. Meanwhile, the study of the damped case has relied on the development of new perturbation methods for the weakly damped scenario \cite {Gulevich}, and the use of conventional or sofisticated computational techniques \cite{Macias-Supra}.

From an experimental perspective, the phenomenon of supratransmission in nonlinear media described by sine-Gordon equations was first observed in mechanical systems of coupled pendula \cite{Geniet-Leon2}, and applications to the design of digital amplifiers of ultra weak signals \cite{Khomeriki-Leon2} and light detectors sensitive to very weak excitations \cite{Chevriaux} have been realized recently. Further applications to optical wave\-guide arrays using the discrete nonlinear Schr\"{o}dinger equation \cite{Khomeriki2}, the realization of light filters \cite{Khomeriki-Ruffo}, and the propagation of binary signals in undamped or weakly damped mechanical chains of oscillators \cite{Macias-Signals} have been recently proposed. 

In this article, we tackle the problem of propagating binary signals in damped, discrete sine-Gordon systems through amplitude modulation (a rather interesting physical problem not studied hitherto), with future studies of signal transmission in Josephson-junction arrays in mind. We make use of analytical approximations and perturbation methods derived from the continuous limiting situation (equivalently, the strongly coupled case) in order to propose the design of a propagation device for binary signals in damped mechanical systems. Our numerical study will be based on a novel computational technique with energy-invariance properties \cite{Macias-Supra}, and the proposed designs will be based on the serial concatenation of such systems.

Section \ref{Sec2} of this work is devoted to introduce the model under study, analytical results on the exact solution of the classical sine-Gordon equation, and the perturbation technique used; the presentation of the computational method to be employed, the finite-difference scheme to estimate the continuous energy and the numerical prediction of the supratransmission occurrence will be the topic of study in Section \ref {Sec2-5}. Section \ref{Sec3} presents the mathematical procedure to generate binary signals, and analyzes characteristics of the propagating signals in the discrete system with respect to the value of the damping coefficients and the normalized bias current. In Section \ref{Sec4}, we propose an alternative for the efficient propagation of binary signals in discrete sine-Gordon systems based on the linear concatenation of semi-infinite chain either bound mechanically of energetically. The results of the simulations for both models are presented in this section, and we summarize our conclusions and propose new directions of future research in a final stage.

\section{Analysis\label{Sec2}}

\subsection{Mathematical model}

Throughout this paper, we assume that $\alpha$, $\beta$, $\gamma$ and $c$ are nonnegative real numbers. Likewise, we consider a system $( u _n ) _{n = 1} ^\infty$ of particles satisfying the mixed-value problem studied in \cite{Macias-Supra}, namely,
\begin{equation}
\begin{array}{c}
\displaystyle {\frac {d ^2 u _n} {d t ^2} - \left( c ^2 +\alpha \frac {d} {d t} \right) \Delta ^2 _x u _n + \beta \frac {d u _n} {d t} + V ^\prime (u _n) = 0, }\\ 
		\begin{array}{rl}
        \begin{array}{l}
            {\rm subject\ to:} \qquad \\ \\ \\
        \end{array}
        \left\{
        \begin{array}{ll}
            u _n (0) = 0, & n \in \mathbb {Z} ^+, \\
            \displaystyle {\frac {d u _n} {d t} (0) = 0}, & n \in \mathbb {Z} ^+, \\
            u _0 (t) = \psi (t), & t \geq 0,
        \end{array}\right.
    \end{array}
\end{array}\label{Eqn:DiscreteMain}
\end{equation}
where $\alpha$ and $\beta$ evidently function as the internal and external damping coefficients, respectively. Here, $\Delta ^2 _x u _n$ is used to denote the spatial second-difference $u _{n + 1} - 2 u _n + u _{n - 1}$ for every $n \in \mathbb {Z} ^+$, the boundary-driving function is given by $\psi (t) = A (t) \cos (\Omega t)$ for every $t \in (0 , + \infty)$, and $V (u _n) = 1 - \cos (u _n) - \gamma u_n$ where, due to its analogy with the model describing long Josephson junctions, the parameter $\gamma$ will be denominated the \emph {normalized bias current}. Notice that the Hamiltonian of the $n$-th site is
\begin{equation}
H _n = \frac {1} {2} \left[ \dot {u} _n ^2 + c ^2 (u _{n + 1} - u _n) ^2 \right] + V (u _n),
\end{equation}
for any differentiable function $V$. After including the potential energy from the coupling between the first two oscillators, the total energy of the system becomes
\begin{equation}
E = \sum _{n = 1} ^\infty H _n + \frac {c ^2} {2} (u _1 - u _0) ^2.
\end{equation}

The next result provides us with a convenient way to compute the instantaneous rate of change of the energy for solutions of (\ref {Eqn:DiscreteMain}) with \emph {square-summable} derivatives in time, that is, solutions $( u _n (t) ) _{n = 0} ^\infty$ for which $\sum \dot {u} _n ^2 (t)$ is convergent for every $t \geq 0$. 

\begin{theorem}[Mac\'{\i}as-D\'{\i}az and Puri \cite {Macias-Supra}]
Let $( u _n (t) ) _{n = 0} ^\infty$ be a solution of system {\rm (\ref {Eqn:DiscreteMain})} such that $( \dot {u} _n (t) ) _{n = 0} ^\infty$ is square-summable at any fixed time $t$. The instantaneous rate of change of the total energy in the system is given by
\begin{equation}
\frac {d E} {d t} = c ^2 (u _0 - u _1) \dot {u} _0 - \beta \sum _{n = 1} ^\infty (\dot {u} _n) ^2 - \alpha \left[ \sum _{n = 1} ^\infty (\dot {u} _n - \dot {u} _{n - 1}) ^2 + (\dot {u} _1 - \dot {u} _0) \dot {u} _0 \right].
\end{equation} \qed
\end{theorem}

\begin{figure}[t]
\centerline{\includegraphics[width=0.8\textwidth]{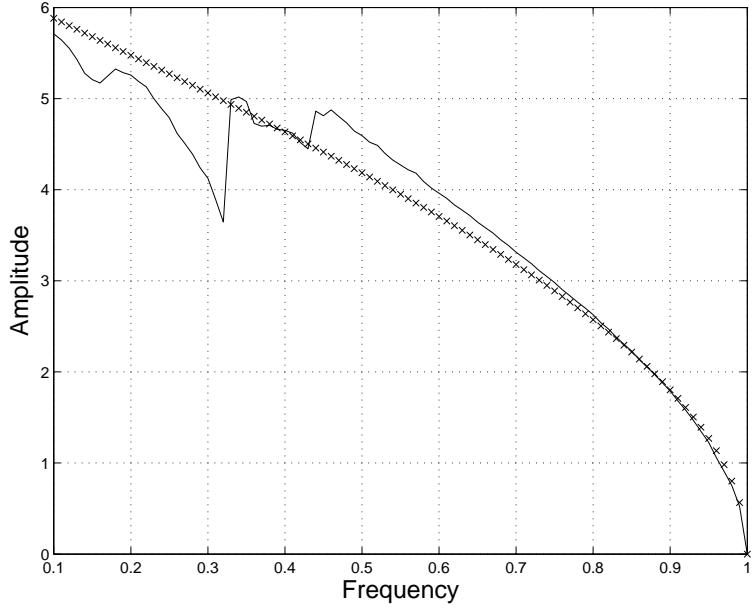}}
\caption{Bifurcation diagram of occurrence of critical amplitude versus driving frequency for problem (\ref {Eqn:DiscreteMain}) with $c = 4$ (solid line). The limiting solution to (\ref {Eqn:ContinuousMain}) is depicted as a sequence of crosses. \label{Fig:Bifurcation}}
\end{figure} 

\begin{corollary} [Geniet and Leon \cite{Geniet-Leon2}]
Let $( u _n (t) ) _{n = 0} ^\infty$ be a solution of undamped system {\rm (\ref {Eqn:DiscreteMain})} with $\gamma = 0$, such that $( \dot {u} _n (t) ) _{n = 0} ^\infty$ is square-summable at any time $t$. Then
\begin{equation}
E (t) = - c ^2 \int _0 ^t \dot {u} _0 (s) (u _1 (s) - u _0 (s)) \ ds
\end{equation} \qed
\end{corollary}

We wish to mention that qualitative results similar to those presented here were obtained using the expressions of the individual energies provided by \cite{Politi}. These formulas were used for validation purposes only.

\subsection{The continuous approximation\label{ContinuousLimit}}

\begin{figure}[t]
\centerline{\includegraphics[width=0.8\textwidth]{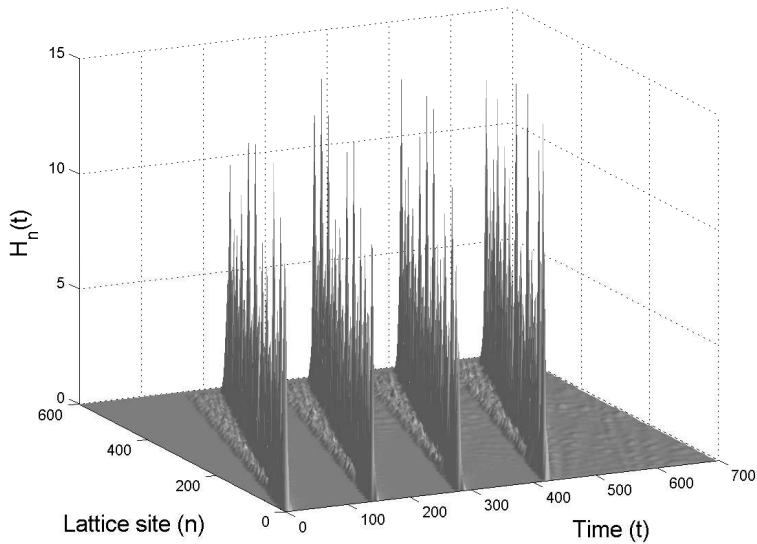}}
\caption{Local energies $H_n (t)$ of unperturbed system (\ref {Eqn:DiscreteMain}) vs. site $n$ and time $t$, corresponding to the transmission of binary signal `$1111$', with $\Omega = 0.9$. \label{Fig:Paper4Fig4}}
\end{figure} 

For theoretical purposes, we will provisionally assume that $\alpha = \beta = 0$, and will consider a long, linear array consisting of $N$ coupled oscillators described by the system of differential equations in (\ref {Eqn:DiscreteMain}), with a coupling coefficient $1 \ll c \ll, N$. Under these conditions, the problem under study can be approximated by the continuous, mixed-value problem  
\begin{equation}
\begin{array}{c}
\displaystyle {\frac {\partial ^2 u} {\partial t ^2} - \frac {\partial ^2 u} {\partial x ^2} + V ^\prime (u) = 0, }\\ 
		\begin{array}{rl}
        \begin{array}{l}
            {\rm subject\ to:} \qquad \\ \\ \\
        \end{array}
        \left\{
        \begin{array}{ll}
            u (x , 0) = 0, & 0 \leq x \leq L, \\
            \displaystyle {\frac {\partial u} {\partial t} (x , 0) = 0}, & 0 \leq x \leq L, \\
            u (0 , t) = \psi (t), & t \geq 0,
        \end{array}\right.
    \end{array}
\end{array}\label{Eqn:ContinuousMain}
\end{equation}
where $L = N / c$, which in turn can be simplified to the study of a sine-Gordon model when the normalized bias current is equal to zero. For a sufficiently long linear chain, a sufficiently short interval of time $[0 , T]$, and a driving function $\psi (t) = A _0 \cos (\Omega t)$, it is reasonable to impose the boundary condition $u _x (L , t) = 0$.

\begin{figure}[t]
\centerline{\includegraphics[width=0.8\textwidth]{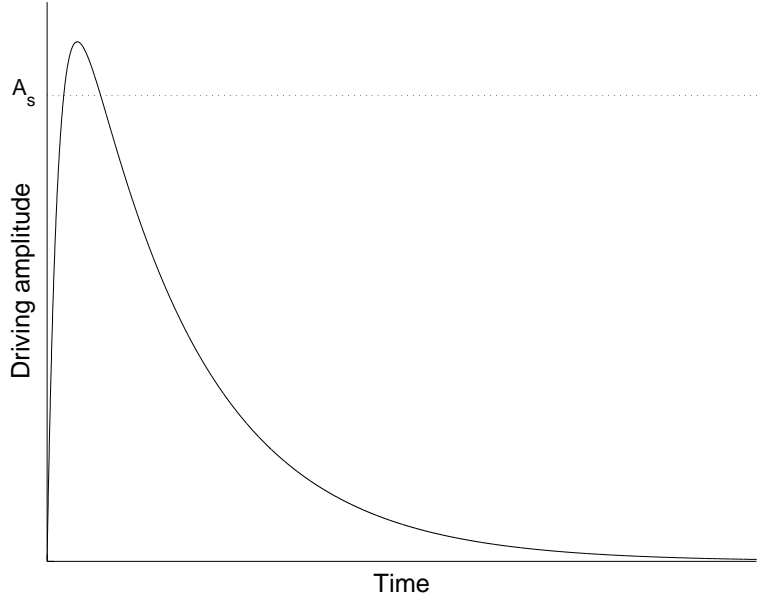}}
\caption{Time-dependent graph of the amplitude function used to generate a bit of `$1$' in a semi-infinite, discrete, mechanical array, with unitary amplification coefficient. \label{FigAbit1}}
\end{figure}

Following \cite{Khomeriki-Leon} now, there exist solutions to problem (\ref {Eqn:ContinuousMain}) in the form 
\begin{equation}
u (x , t) = 4 \arctan [X (x) T (t)],
\end{equation}
with $X (L) = A _L$ and normalized time scale $T$ satisfying $T (t _0) = 1$ for some time $t _0$. We assume that the solutions are periodic in time with period equal to the driving period, and that the solutions adapt to the driving amplitude, that is, $A _0 = 4 \arctan (a)$, which yields $X (0) = a$. These solutions are seen to depend on a free parameter $\lambda$ and the parameter
\begin{equation}
\Gamma = A _L ^{- 2} + \frac {1} {\lambda (1 + A _L ^2)}.
\end{equation}
The relations between $a$, $A _L$ and $\lambda$ are summarized in the following result. Here $\mathbb {K} (m)$, $\sn ( \cdot , m )$, $\cn ( \cdot , m )$ and $\dn ( \cdot , m )$ are the complete elliptic integral of the first kind, the sine-amplitude and the cosine-amplitude Jacobi elliptic function of modulus $m$, and the identity-amplitude Jacobi elliptic function of modulus $m$, respectively.

\begin{figure}[t]
\centerline{\includegraphics[width=0.8\textwidth]{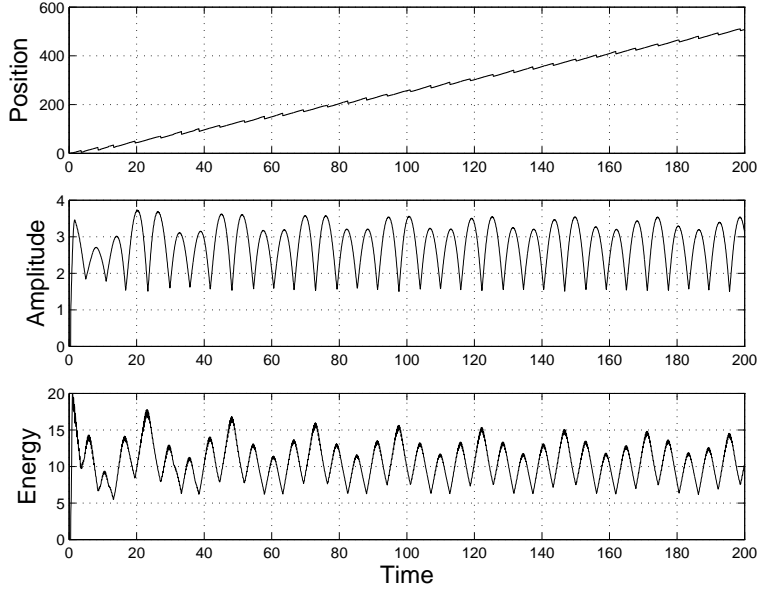}}
\caption{Time-dependent graphs of the position and maximum energy amplitude of the breather generated in unperturbed system (\ref {Eqn:DiscreteMain}) by bit `$1$', with $\Omega = 0.9$. \label{Fig:Paper4Fig5}}
\end{figure}

\begin{theorem} [Khomeriki and Leon \cite {Khomeriki-Leon}] \label{Thm:1} \ 
\begin{enumerate}
\item[{\rm (1)}] If $\lambda > 0$ then $X = A _L \cn (k (x - L) , \mu)$ and $T = \cn (\omega (t - t _0) , \nu)$, with equations $\Omega \mathbb {K} (\nu) = \frac {\pi} {2} \omega$ and $a = A _L \cn (k L , \mu)$ being satisfied for
\begin{equation}
\begin{array}{rclcrcl}
\omega ^2 & = & \lambda (1 + \Gamma), & & \nu ^2 & = & \displaystyle {\frac {1} {1 + \Gamma}}, \\
k ^2 & = & \lambda \Gamma \left( A _L ^2 + \displaystyle {\frac {1} {\Gamma A _L ^2}} \right), & & \mu ^2 & = & \displaystyle {\frac {\Gamma A _L ^2} {1 + \Gamma A _L ^4}}.
\end{array}
\end{equation}
\item[{\rm (2)}] If $\lambda < 0$ and $\Gamma A _L ^4 < -1$ then $X = A _L \dn (k (x - L) , \mu)$ and $T = \sn (\omega (t - t _1) , \mu)$, with equations $\Omega \mathbb {K} (\nu) = \frac {\pi} {2} \omega$ and $a = A _L \dn (k L , \mu)$ satisfied for
\begin{equation}
\begin{array}{rclcrcl}
\omega ^2 & = & \lambda \Gamma, & & \nu ^2 & = & - \displaystyle {\frac {1} {\Gamma}}, \\ 
k ^2 & = & \lambda \Gamma A _L ^2, & & \mu ^2 & = & 1 + \displaystyle {\frac {1} {\Gamma A _L ^4}}, \\
t _1 & = & \displaystyle {t _0 + \frac {\mathbb {K} (\nu)} {\omega}.}
\end{array}
\end{equation}
\item[{\rm (3)}] If $\lambda < 0$ and $\Gamma A _L ^4 > - 1$ then $X = A _L \dn ^{- 1} (k (x - L) , \mu)$ and $T = \sn (\omega (t - t _1) , \mu)$, where $\Omega \mathbb {K} (\nu) = \frac {\pi} {2} \omega$ and $a = A _L \dn ^{- 1} (k L , \mu)$ are satisfied for 
\begin{equation}
\begin{array}{rclcrcl}
k ^2 & = & - \displaystyle {\frac {\lambda} {A _L ^2}}, & & \mu ^2 & = & 1 + \Gamma A _L ^4,
\end{array}
\end{equation}
and for $\omega ^2$, $\nu ^2$ and $t _1$ as in $(2)$. \qed
\end{enumerate}
\end{theorem}

\begin{corollary} [Khomeriki and Leon \cite {Khomeriki-Leon}] \label{Coro}
The limiting solution of mixed-value problem {\rm (\ref {Eqn:ContinuousMain})} as $L \rightarrow \infty$ is given by
\begin{equation}
u (x , t) = 4 \arctan \left( a _s \frac {\sin ( \omega (t - t _0) )} {\cosh (k x)} \right),
\end{equation}
when $\psi (t) = A _0 \sin (\Omega t)$. Here $a _s ^2 = A ^2 (1 - \mu ^2) ^{- 1}$ and the rest of the parameters are as in $(3)$ of Theorem $\ref {Thm:1}$. \qed
\end{corollary}

\subsection{Perturbation analysis\label{PertAnal}}

\begin{figure}[t]
\centerline{\includegraphics[width=0.8\textwidth]{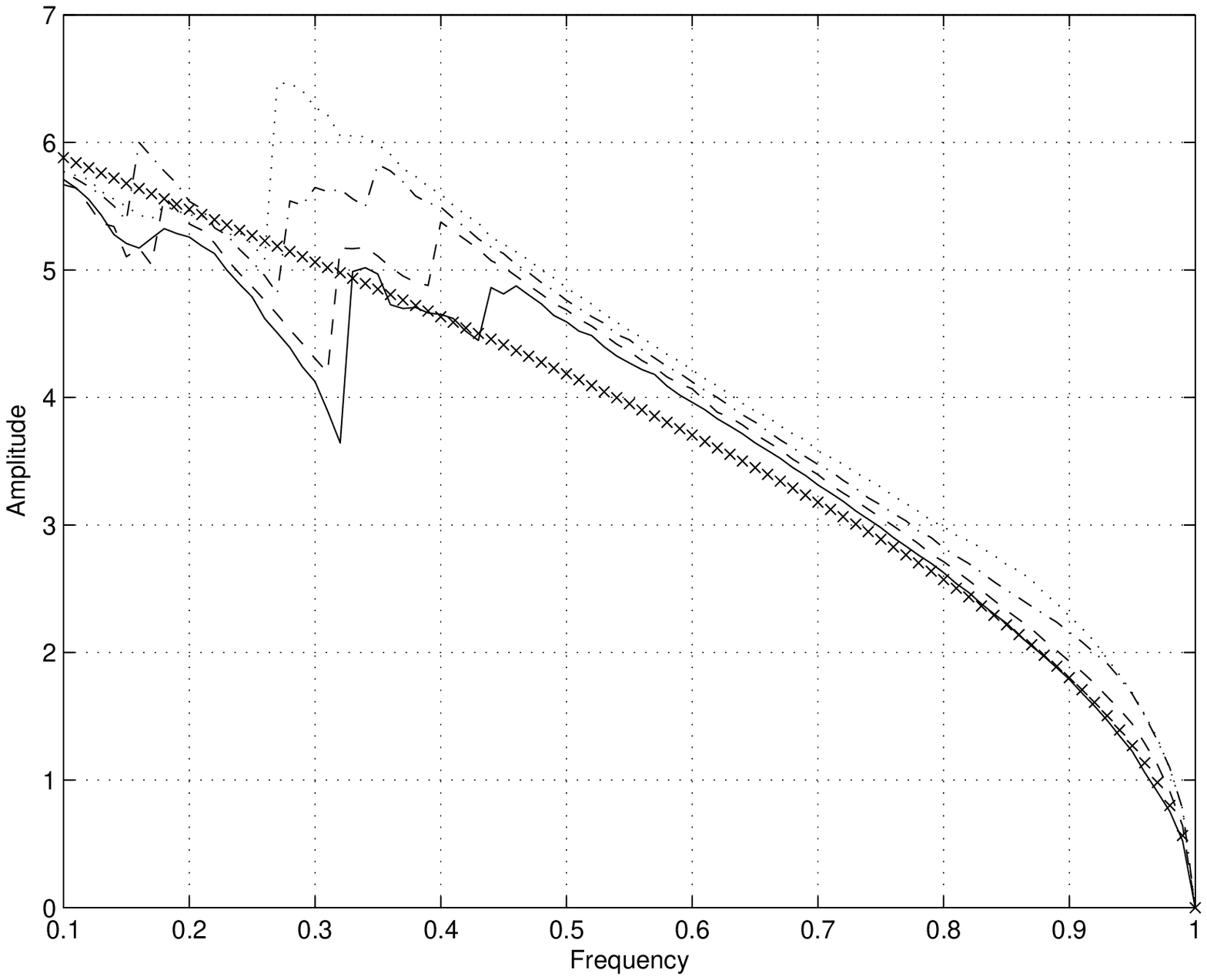}}
\caption{Bifurcation diagram of occurrence of critical amplitude versus driving frequency for problem (\ref {Eqn:DiscreteMain}) with $c = 4$, for $\beta = 0$ (solid), $0.1$ (dashed), $0.2$ (dashed-dotted), $0.3$ (dotted). The limiting solution to unperturbed system (\ref {Eqn:ContinuousMain}) is depicted as a sequence of crosses. \label{Fig:BifurcationAlpha}}
\end{figure} 

In this section, we will assume that $u (x , t)$ represents an exact solution of the unperturbed sine-Gordon equation in (\ref {Eqn:ContinuousMain}) with $- \infty < x < \infty$, and that $\beta$ and $\gamma$ are small, nonnegative real numbers. We will consider the perturbed sine-Gordon model
\begin{equation}
\frac {\partial ^2 v} {\partial t ^2} - \frac {\partial ^2 v} {\partial x ^2} + \beta \frac {d v} {d t} + \sin v - \gamma = 0, \qquad - \infty < x < \infty.
\end{equation}

Following \cite {Gulevich}, we assume the existence of a solution of the perturbed sine-Gordon equation of the form $u( g (\phi (t)) x , T (t) , \phi (t))$ with 
\begin{equation}
T (t) = \int _0 ^t g (\phi (s)) \phi (s) ds \qquad {\rm and} \qquad g (\phi) = (1 \pm \phi ^2) ^{- 1 / 2},
\end{equation}
for some suitable function $\phi$. If we let $u$ be a localized intrinsic mode (static breather) solution of the unperturbed model confined in some region $| x | < \delta$, then 
\begin{equation}
\frac {\partial u} {\partial t} \approx \frac {\partial u} {\partial T} g (\phi) \phi,
\end{equation}
and it follows that the Hamiltonian of the unperturbed sine-Gordon equation can be approximated through the expression
\begin{equation}
H ^* (\phi (t)) = \int _{- \infty} ^\infty \left[ \frac {1} {2} \left( \frac {\partial u} {\partial T} g (\phi) \phi \right) ^2 + \frac {1} {2} \left( \displaystyle {\frac {\partial u} {\partial x}} \right) ^2 + 1 - \cos u \right] dx.
\end{equation}
It can be shown then that the relation between the variables $\phi$, $t$ and $T$ is described through the system 
\begin{equation}
\left\{
\begin{array}{rcl}
\displaystyle {\frac {d \phi} {d T}} & = & - \displaystyle {\frac {1} {\phi g (\phi) \displaystyle {\frac {d H ^*} {d \phi}}}} \displaystyle{ \int _{- \infty} ^\infty \left[ \beta \left( \displaystyle {\frac {d u} {d t}} \right) ^2 + \gamma \frac {d u} {d t} \right] dx}, \\
\displaystyle {\frac {d t} {d T}} & = & \displaystyle {\frac {1} {\phi g (\phi)}}.
\end{array}
\right.
\end{equation}

\section{Computational approach\label{Sec2-5}}

\subsection{Finite-difference schemes\label{NumScheme}}

\begin{figure}[t]
\centerline{\includegraphics[width=0.8\textwidth]{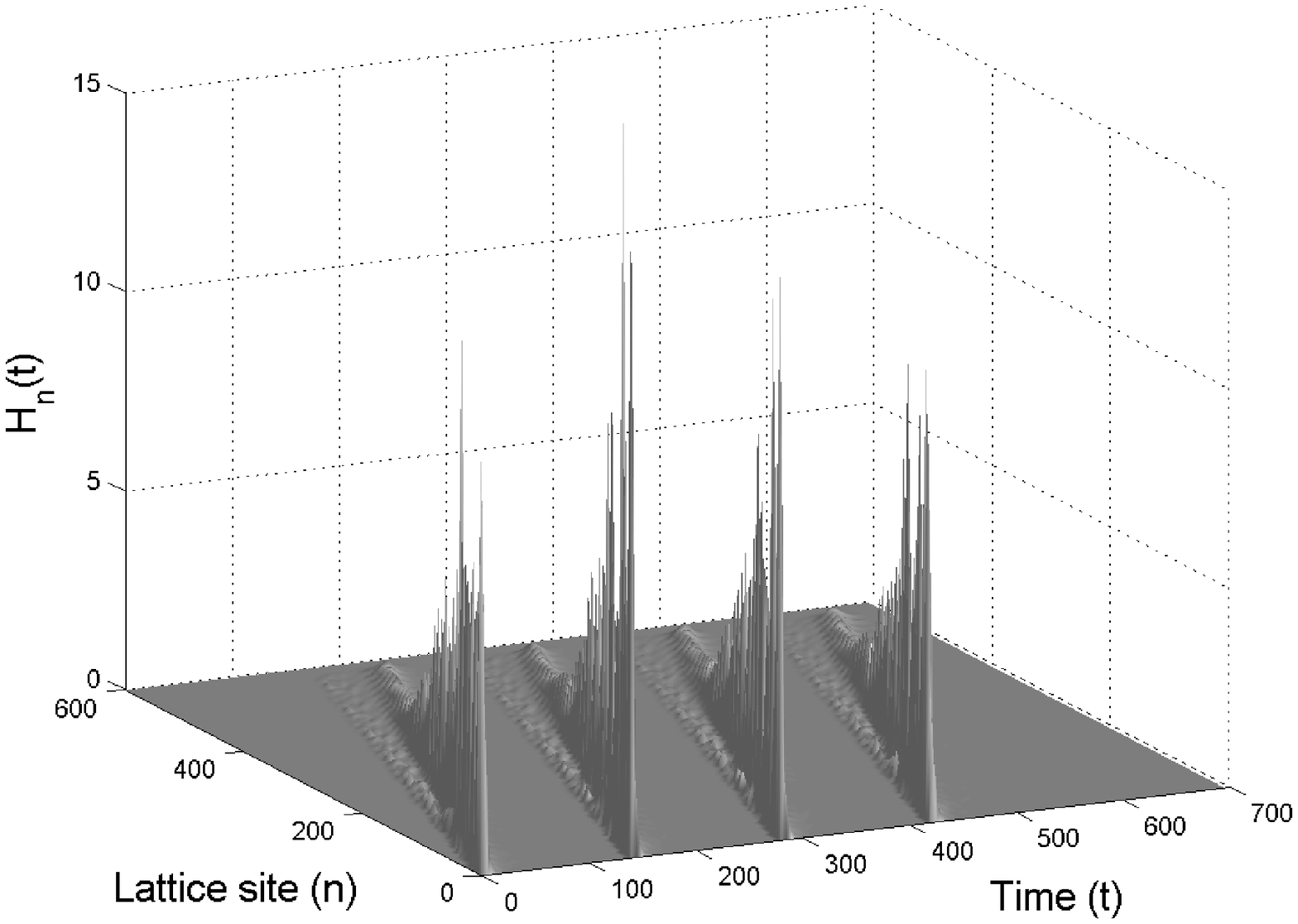}}
\caption{Local energies $H_n (t)$ of perturbed system (\ref {Eqn:DiscreteMain}) with $\beta = 0.01$ vs. lattice site $n$ and time $t$, corresponding to the transmission of binary signal `$1111$', with $\Omega = 0.9$. \label{Fig:Paper4Fig6}}
\end{figure} 

We consider a finite system of $N$ differential equations satisfying (\ref {Eqn:DiscreteMain}), and a regular partition $0 = t _0 < t _1 < \dots < t _M = T$ of the time interval $[0 , T]$ with time step equal to $\Delta t$. For each $k = 0 , 1 , \dots , M$, let us represent the approximate solution to our problem on the $n$-th lattice site at time $t _k$ by $u ^k _n$. If we convey that $\delta _t u _n ^k = u _n ^{k + 1} - u _n ^{k - 1}$, that $\delta ^2 _t u _n ^k = u _n ^{k + 1} - 2 u _n ^k + u _n ^{k - 1}$ and that $\delta ^2 _x u _n ^k = u _{n + 1} ^k - 2 u _n ^k + u _{n - 1} ^k$, the $N$ differential equations in our problem take then the discrete form
\begin{equation}
\displaystyle {\frac {\delta ^2 _t u _n ^k} {(\Delta t) ^2} - \left( c ^2 + \frac {\alpha} {2 \Delta t} \delta _t \right) \delta ^2 _x u _n ^k + \frac {\beta ^\prime} {2 \Delta t} \delta _t u _n ^k + \frac {V (u _n ^{k + 1}) - V (u _n ^{k - 1})} {u _n ^{k + 1} - u _n ^{k - 1}}} = 0,
\label{Eqn:DiscreteMainDiscr}
\end{equation}
where $\beta ^\prime$ includes both the effect of external damping and a simulation of an absorbing boundary slowly increasing in magnitude on the last $N - N _0$ oscillators. More concretely, we let $u _{N + 1} = u _N$ at all time, and let $\beta ^\prime$ be the sum of external damping and the function 
\begin{equation}
\beta ^{\prime \prime} (n) = 0.5 \left[ 1 + \tanh \left( \displaystyle {\frac {2 n - N _0 + N} {6}} \right) \right],
\end{equation}
where usually $N _0 = 50$ and $N \geq 200$.

Finite-difference scheme (\ref {Eqn:DiscreteMainDiscr}) (which is a modified version of the one developed in \cite{Macias-Puri} to compute radially symmetric solutions of modified Klein-Gordon equations) is consistent with our mixed-value problem and conditionally stable, having the inequality $\left(c \Delta t \right) ^2 < 1 + \left(\alpha + \beta ^\prime / 4\right) \Delta t$ as a necessary condition for stability when $\beta ^\prime$ is assumed constant \cite{Macias-Supra}. Moreover, if the energy of the system at the $k$-th time step is computed using the expression 
\begin{equation}
E _k = \sum _{n = 1} ^ M H _n ^k + \frac {c ^2} {2} (u _1 ^{k + 1} - u _0 ^{k + 1}) (u _1 ^k - u _0 ^k),
\end{equation}
where the individual energy of the $n$-th lattice site is computed by
\begin{eqnarray}
H _n ^k & = & \frac {1} {2} \left( \frac {u _n ^{k + 1} - u _n ^k} {\Delta t} \right) ^2 + \frac {c ^2} {2} (u _{n + 1} ^{k + 1} - u _n ^{k + 1}) (u _{n + 1} ^k - u _n ^k) \\ \nonumber
 & & \qquad\quad + \frac {V (u _n ^{k + 1}) + V (u _n ^k)} {2},
\end{eqnarray}
then the discrete rate of change of energy turns out to be a consistent approximation of order $\mathcal {O} (\Delta t) ^2$ for the corresponding instantaneous rate of change, while the approximations of the individual and the total energies are consistent estimates order $\mathcal {O} (\Delta t)$ of their continuous counterparts.

\begin{figure}[t]
\centerline{\includegraphics[width=0.8\textwidth]{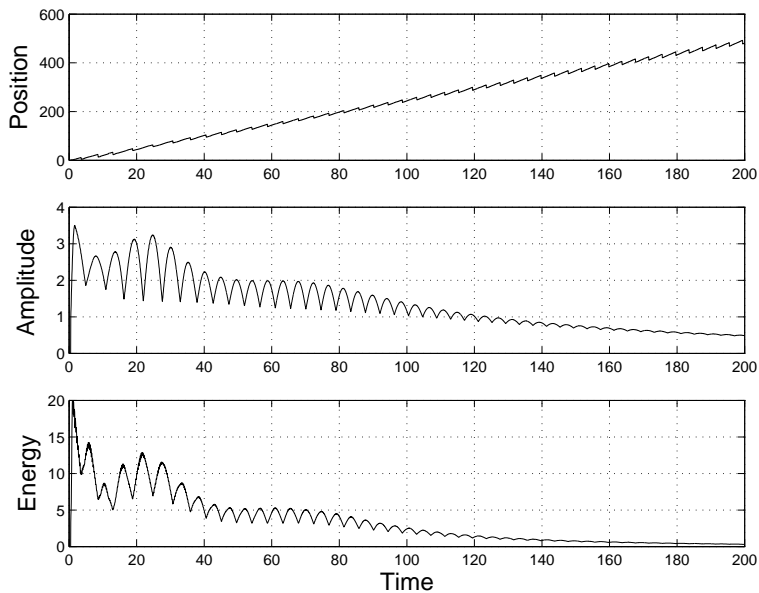}}
\caption{Time-dependent graphs of the position and maximum energy amplitude of the breather generated by bit `$1$' in (\ref {Eqn:DiscreteMain}) with $\beta = 0.01$ and $\Omega = 0.9$. \label{Fig:Paper4Fig7}}
\end{figure}

\subsection{Supratransmission threshold}

It has been proved that a discrete or continuous medium described by damped problem (\ref {Eqn:DiscreteMain}) or undamped problem (\ref {Eqn:ContinuousMain}), respectively, can undergo nonlinear energy transmission when the driving amplitude is increased above a critical threshold that depends on the driving frequency, and which has been numerically predicted in \cite{Macias-Supra,Khomeriki-Leon}. In the discrete case particularly, the numerical results obtained using the computational technique described in the previous section have produced bifurcation diagrams for various values of $\alpha$, $\beta$ and $\gamma$, which are in excellent agreement with the undamped, unbiased formulation of the problem for large coupling coefficients. In this case the bifurcation threshold can be approximated via the continuous approximation provided by Corollary \ref{Coro}, which in turn can be approximated through
\begin{equation}
A _s = 4 \arctan \left(\frac {\sqrt {1 - \Omega ^2}} {\Omega} \right). \label {Eqn:TheoApprox}
\end{equation}

For the sake of convenience, we have included a bifurcation diagram in Figure \ref {Fig:Bifurcation}, which shows the numerical (solid line) and theoretical (sequence of crosses) critical amplitude at which supratransmission starts for each frequency $\Omega$ in the forbidden band gap. The theoretical approximation was realized using (\ref {Eqn:TheoApprox}).

In the continuous limit, the mechanism of the bifurcation is explained by the fact that the medium starts to transmit energy in the form of moving breathers once the driving amplitude reaches its critical value \cite {Khomeriki-Leon}, and the medium continues transmitting energy well until a vanishing driving amplitude is reached. In the discrete case, it has been checked numerically that this phenomenon still occurs \cite {Geniet-Leon}, and that the expression of the moving breathers is given by
\begin{equation}
u _n (t) = 4 \arctan \left[ \frac {r \sin \left( \frac {1} {\sqrt {1 + r ^2}} \frac {t - v n / c} {\sqrt {1 - v ^2}} \right)} {\cosh \left( \frac {r} {\sqrt 1 + r ^2} \frac {n / c - v t} {\sqrt {1 - v ^2}} \right) } \right],
\end{equation}
where $r$ and $v$ represent, respectively, the group amplitude and the group velocity of the breather. Numerical experiments have shown that $r$ and $v$ are variables that linearly depend on $A _s$, for values of $0.87 < \Omega < 1$ (see \cite {Geniet-Leon2}). Moreover, we have verified numerically that the amplitude of the emitted breathers tend to decrease as the driving amplitude is decreased from $A _s$ toward zero. 

\begin{figure}[t]
\centerline{\includegraphics[width=0.8\textwidth]{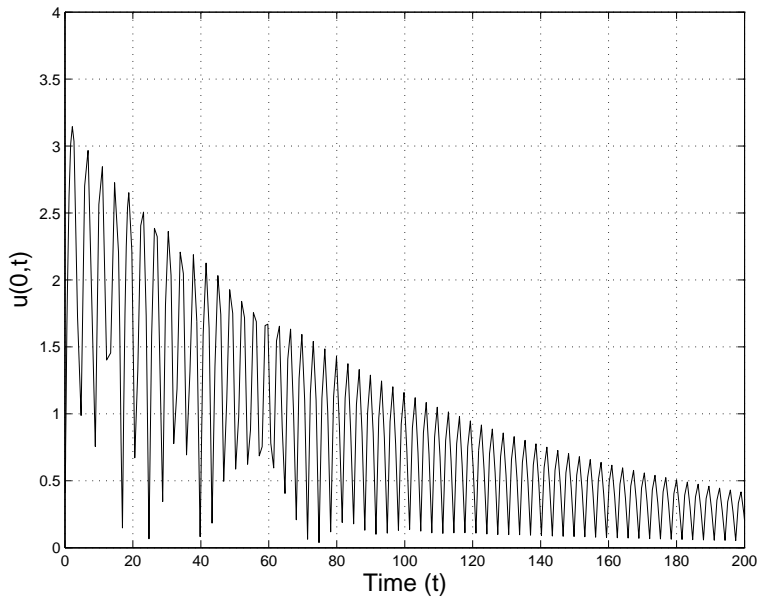}}
\caption{Time-dependent graph of decay of a static breather in infinite system (\ref {Eqn:ContinuousMain}), for constant damping $\beta = 0.01$. \label{Fig:AmplitudeOrigin}}
\end{figure}

\section{Numerical results\label{Sec3}}

Following \cite{Macias-Signals}, in this section we consider a semi-infinite, mechanical system described by (\ref {Eqn:DiscreteMain}). A single binary bit $b$ will be transmitted into the medium during a fixed and sufficiently long period of time $P$ equal to an integer multiple of the driving period, by letting
\begin{equation}
A (t) = \frac {8} {5} b C A_s \left(e ^{- \Omega t / 4.5} - e ^{- \Omega t / 0.45} \right), \label{Eqn:Amplitude}
\end{equation}
for every $t \in [0 , P]$, where $A _s$ represents the amplitude threshold at which nonlinear supratransmission begins, and $C$ is an \emph {amplification constant} that depends of $\Omega$. It is worth noticing that $A$ is identically equal to zero for a bit $b$ equal to zero; when $b$ is equal to one, the qualitative behavior of $A$ over a period equal to several times the driving period is presented in Figure \ref{FigAbit1}. 

\begin{figure}[t]
\centerline{\includegraphics[width=0.8\textwidth]{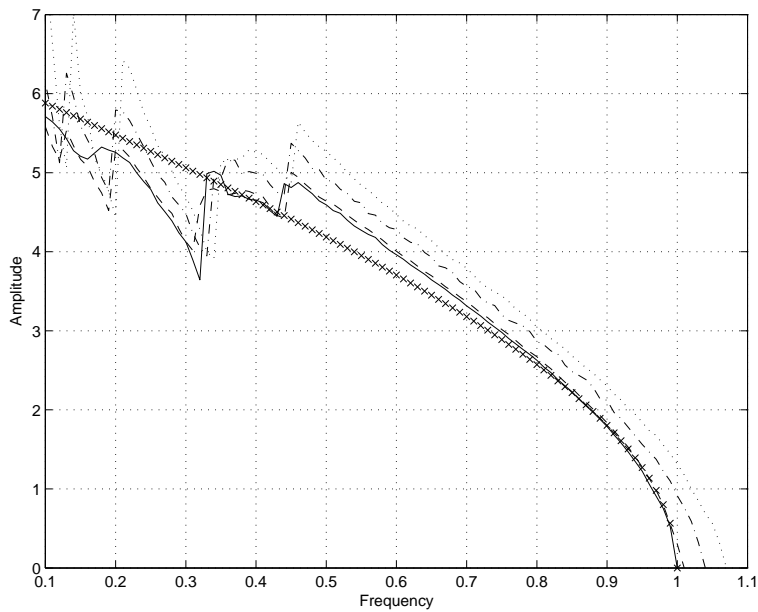}}
\caption{Bifurcation diagram of occurrence of critical amplitude versus driving frequency for problem (\ref {Eqn:DiscreteMain}) with $c = 4$, for $\alpha = 0$ (solid), $0.1$ (dashed), $0.4$ (dashed-dotted), $0.7$ (dotted). The limiting solution to unperturbed system (\ref {Eqn:ContinuousMain}) is depicted as a sequence of crosses. \label{Fig:BifurcationBeta}}
\end{figure} 

Before we study the effects of damping and the normalized current in the moving breathers generated by the amplitude function above, it is highly convenient to introduce some terminology. Thus, for a breather moving away from the irradiating source at a constant velocity $v$, we assume the existence of constants $\delta > 0$ and $n _0 \in \mathbb {Z}$ with the property that the breather is approximately equal to zero outside of the region $A$ defined by $|n - (n _0 + v t)| \leq \delta$. In such case, we say that the \emph {position of the breather} at time $t$ is equal to $n _0 + v t$, and that its \emph {amplitude} and \emph {energy} are, respectively, 
\begin{equation}
A _b (t) = \displaystyle {\max _{n \in A}}\ | u _n (t) | \qquad {\rm and} \qquad
E _b (t) = \displaystyle {\max _{n \in A}}\ H _n (t).
\end{equation}
Of course, this definition of position of a breather is rather subjective. A bit more formally, if we know that the breather is located in a region $R$ at time $t$, its position may be defined as the integer $n _1 \in R$ such that $H _{n _1} (t) = \displaystyle {\max _{n \in R}}\ H _n (t)$.

\subsection{The ideal case}

Let us consider an undamped, mixed-value problem (\ref {Eqn:DiscreteMain}) for which the normalized bias current is equal to zero, the coupling coefficient is equal to $4$ and the period of bit generation equals $20$ driving periods. Numerically, we consider a finite chain system of length $N = 600$, a time step of $0.05$, and use the computational scheme sketched in Section \ref{NumScheme}. A bifurcation diagram similar to the one found in \cite{Macias-Supra} is readily obtained, and the critical amplitudes $A _s$ are used in the determination of the velocities for single moving breathers. Under this situation, a binary signal of `$1111$' is transmitted into chain (\ref {Eqn:DiscreteMain}) driven at a frequency equal to $0.9$, with an amplification constant equal to $A_s$.

Figure \ref{Fig:Paper4Fig4} shows the development of the local energy in this system with respect to time and lattice site. The fact that moving breathers are generated in the chain is clear from the graph. The fact that the first of the breathers moves at approximately a constant speed is shown by the upper graph in Figure \ref {Fig:Paper4Fig5}, while its maximum local energy is analyzed in the bottom portion of the same figure. Clearly, this maximum is above a positive cutoff limit, say, equal to $5$.

\begin{figure}[t]
\centerline{\includegraphics[width=0.8\textwidth]{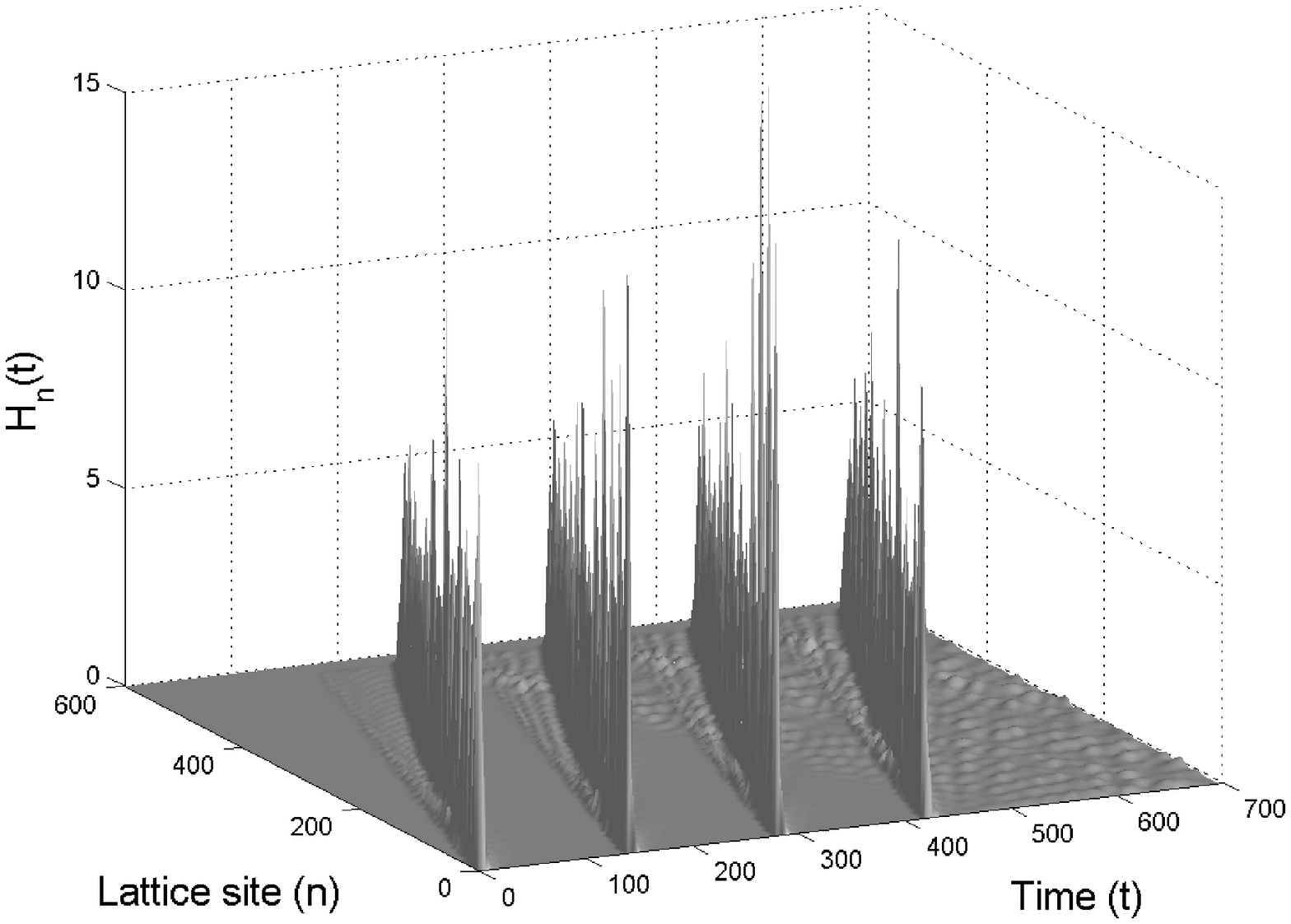}}
\caption{Local energies $H_n (t)$ of (\ref {Eqn:DiscreteMain}) with $\alpha = 0.05$ vs. lattice site $n$ and time $t$, corresponding to the transmission of binary signal `$1111$', with $\Omega = 0.9$. \label{Fig:Paper4Fig8}} 
\end{figure}

In order to find a more reliable estimate of the lower cutoff limit for the maximum local energy carried by a breather generated using (\ref {Eqn:Amplitude}), we have transmitted a single bit `$1$' into a chain consisting of $6000$ lattice sites during a period of time equal to $2000$, with all other parameters as in the previous paragraph. As a result, we obtain that the absolute minimum and maximum values of the maximum local energy in a single breather are equal to $5.6965$ and $17.9835$, respectively, whence it particularly follows that a lower cutoff limit of $5$ would actually prove to be adequate.

\subsection{External damping\label{ExtDamp}}

The next step in our investigation will be to describe the effects of internal and external damping, and the normalized bias current on the behavior of propagating signals in (\ref {Eqn:DiscreteMain}). More concretely, we wish to establish the qualitative effects of these parameters on the local energy amplitude and speed evolution of the breather solutions with respect to lattice site. To that effect, we consider the same mixed-value problem studied in the previous section with similar choices of parameters, and provide a bifurcation diagram for several values of $\beta$ in Figure \ref {Fig:BifurcationAlpha}.

\begin{figure}[t]
\centerline{\includegraphics[width=0.8\textwidth]{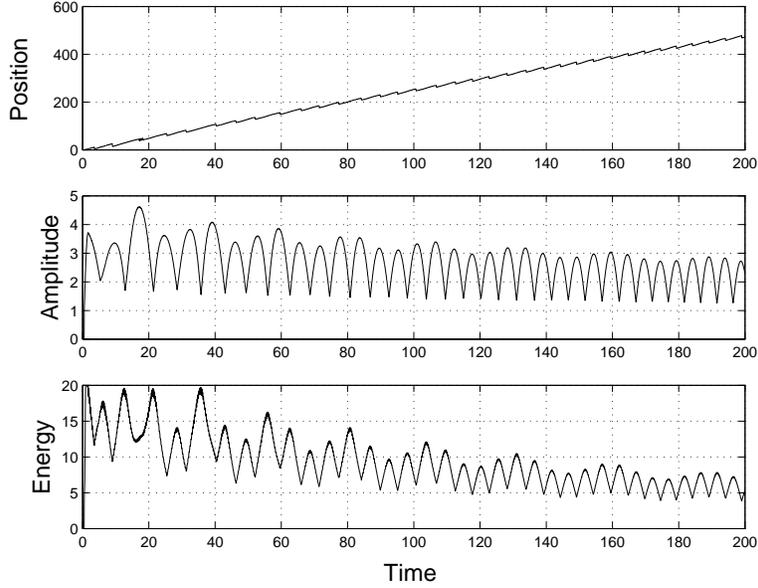}}
\caption{Time-dependent graphs of the position and maximum energy amplitude of the breather generated by bit `$1$' in (\ref {Eqn:DiscreteMain}) with $\alpha = 0.05$. \label{Fig:Paper4Fig9}}
\end{figure}

The binary signal `$1111$' will be propagated again using the driving amplitude function (\ref {Eqn:Amplitude}) assuming that the medium is externally damped with a coefficient of $0.01$. In this context, Figure \ref {Fig:Paper4Fig6} shows the time and lattice site evolution of the local energy carried by the breathers generated by the transmitted signals. These results evidence a drastic decay in the amplitudes of the local energies of a breather as it moves in the chain, a claim that is confirmed by Figure \ref {Fig:Paper4Fig7}, in which we have tracked the position and the development of the maximum local energy of the moving breather generated by a single bit `$1$'. 

Two conclusions are now evident from the graphs in Figure \ref {Fig:Paper4Fig7}: first, the breathers generated using our transmission process move away from the source at a constant speed; second, a decrease of the maximum local energy amplitude carried by the breather with respect to time (and, hence, with respect to lattice site) is now apparent. 

From Section \ref {ContinuousLimit}, the solution to problem (\ref {Eqn:ContinuousMain}) in the continuous limit case is given by
\begin{equation}
u (x , t) = 4 \arctan \left( \frac {\lambda c \sin (\Omega t)} {\Omega \cosh (\lambda x)} \right), \label{Formula}
\end{equation}
where $\lambda = \arccosh \left( 1 + (1 - \Omega ^2) / 2 c ^2 \right)$. Using this solution and applying the perturbation method of Section \ref{PertAnal} with $g (\phi) = 1 / \sqrt {1 + \phi ^2}$, we obtain the time evolution of the amplitude at the origin of a static breather initially described by (\ref {Formula}), with an external damping coefficient equal to $0.01$. The results, which are presented in Figure \ref {Fig:AmplitudeOrigin}, evidence a good agreement with the time evolution of the amplitude decay of the breather presented in Figure \ref {Fig:Paper4Fig6}.

\subsection{Internal damping}

\begin{figure}[t]
\centerline{\includegraphics[width=0.8\textwidth]{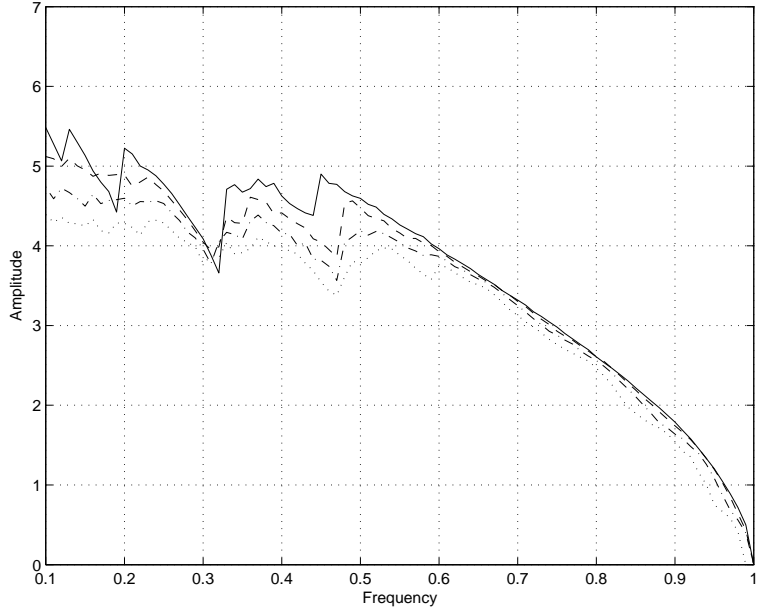}}
\caption{Bifurcation diagram of occurrence of critical amplitude versus driving frequency for problem (\ref {Eqn:DiscreteMain}) with $c = 4$, for $\gamma = 0$ (solid), $0.01$ (dashed), $0.02$ (dashed-dotted), $0.03$ (dotted). \label{Fig:BifurcationGamma}}
\end{figure} 

Consider again problem (\ref {Eqn:DiscreteMain}) with the same parameters as in Section \ref{ExtDamp} and similar numerical setting, describing a medium with no external damping and internal damping coefficient equal to $0.05$. Again, the binary signal `$1111$' will be propagated into the medium using driving amplitude function (\ref {Eqn:Amplitude}). For convenience, Figure \ref {Fig:BifurcationBeta} provides a bifurcation diagram of energy transmission for several values of $\alpha$. 

Under this state of matters, Figure \ref{Fig:Paper4Fig8} presents the dependence of the local energy amplitude of each lattice site with respect to time. The graphs shows a decrease in time of the local energy amplitude of the generated breathers. To verify this claim, Figure \ref{Fig:Paper4Fig9} presents the time development of the position and maximum local energy attained by the first breather produced. This last figure also suggests that the breathers generated in this case move away from the driving boundary with approximately a constant speed, a fact that was corroborated using linear regression analysis.

We must remark that, like in the case of external damping, higher values of internal damping produce graphs of amplitude and local energy that decrease faster toward zero in time, as it is expected. In any case, we have verified that the phase velocity of the emitted breathers remains constant.

\subsection{Normalized bias current\label{Bias}}

Finally, we carry out a similar analysis for the normalized bias current. 

\begin{figure}[t]
\centerline{\includegraphics[width=0.8\textwidth]{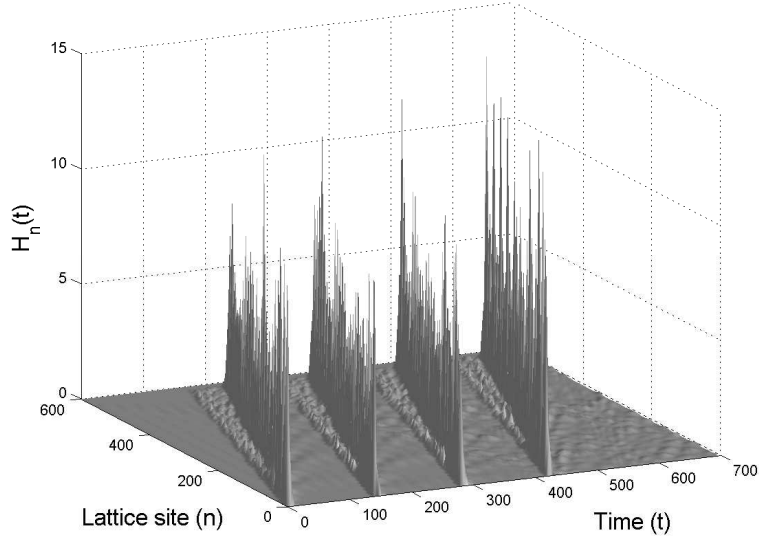}}
\caption{Local energies $H_n (t)$ of (\ref {Eqn:DiscreteMain}) with $\gamma = 0.1$ vs. lattice site $n$ and time $t$, corresponding to the transmission of binary signal `$1111$', with $\Omega = 0.9$. \label{Fig:Paper4Fig10}} 
\end{figure}

\begin{figure}[t]
\centerline{\includegraphics[width=0.8\textwidth]{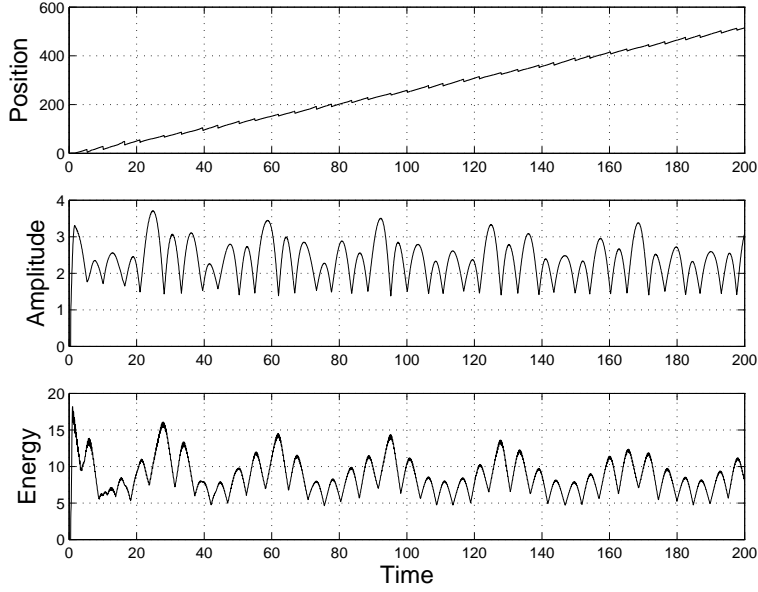}}
\caption{Time-dependent graphs of the position and maximum energy amplitude of the breather generated by bit `$1$' in (\ref {Eqn:DiscreteMain}) with $\gamma = 0.1$ and $\Omega = 0.9$. \label{Fig:Paper4Fig11}}
\end{figure}

Figure \ref{Fig:BifurcationGamma} shows the effect of several values of the parameter $\gamma$ in the occurrence of the supratransmission threshold. As before, the binary signal `$1111$' is generated by the driving boundary of a mechanical chain with parameters as in the previous section, and $\gamma = 0.1$. Figure \ref {Fig:Paper4Fig10} shows then the motion of the generated breathers in our mechanical system, while Figure \ref {Fig:Paper4Fig11} provides time-dependent graphs of velocity, amplitude and local energy of the breather generated by the first bit. For this choice of $\gamma$, a slight decrease in the amplitude and the local energy of the breathers is shown.

\section{Application\label{Sec4}}

\subsection{Concatenated mechanical chains}

Let $\ell$ be a positive integer, and for every $j = 0 , 1, \dots \ell$, assume that $\Omega _j$ and $c _j$ are positive numbers, that $\alpha _j$, $\beta _j$ and $\gamma _j$ are nonnegative real numbers, and that $L _j$ is a positive integer. Suppose that the sequence $( \mathbf {u} _n ) _{n = 1} ^\infty$ of vectors $\mathbf {u} _n = (u _{j n}) _{j = 1} ^\ell$ satisfies the system of concatenated equations
\begin{equation}
\left\{
\begin{array}{c}
\begin{array}{rcl}
\displaystyle {\frac {d ^2 {u} _{j n}} {d t ^2} - \left( c _j ^2 + \alpha _j \frac {d} {d t} \right) \Delta ^2 _x {u} _{j n} + \beta _j \frac {d {u} _{j n}} {d t} + V ^\prime ( {u} _{j n} )} & = & 0,
\end{array}\\ 
		\begin{array}{rl}
        \begin{array}{l}
            {\rm subject\ to:} \qquad \\ \\ \\
        \end{array}
        \left\{
        \begin{array}{ll}
            {u} _{j n} (0) = 0, & n \in \mathbb {Z} ^+, \\
            \displaystyle {\frac {d  {u} _{j n}} {d t} (0) = 0}, & n \in \mathbb {Z} ^+, \\
            {u} _{j 0} (t) = \psi _j (t), & t \geq 0,
        \end{array}\right.
    \end{array}
\\
    {\rm for} \ j = 0 , 1 , \dots , \ell, \qquad\qquad\qquad\qquad\qquad\qquad\qquad
\end{array} \label{Eqn:ApplicationEq}
\right. 
\end{equation}
with potential functions given by $V _j (u) = 1 - \cos (u) - \gamma _j u$, and boundary-driving functions $\psi _j (t) = A _j (t) \cos(\Omega _j t)$ for every $t \in (0 , + \infty)$ and $j = 0 , 1 , \dots , \ell$, where each frequency $\Omega _j$ is assumed to be in the forbidden band gap of the $j$-th discrete array. The number $\ell$ will be called the \emph {length of the system} of concatenated arrays.

Let $P _0$ be equal to a multiple of the period of the harmonic driving in the zeroth mechanical array. As in Section \ref{Sec3}, a single bit of information $b$ will be transmitted during a period of time equal to $P _0$ by modulating the amplitude of the driving oscillator in the zeroth array through
\begin{equation}
A _0 (t) = \left\{
\begin{array}{ll}
\displaystyle {\frac {8} {5} b C A_s \left(e ^{- \Omega _0 t / 4.5} - e ^{- \Omega _0 t / 0.45} \right),} & {\rm if} \ b = 1. \\ \\
0, & {\rm if} \ b = 0.
\end{array}
\right. \label{Eqn:AmplitudeMult}
\end{equation}
The reception device of the first mechanical array will be initialized then at time $L _0 / v$, where $v$ represents the phase velocity of the generated breathers in the zeroth array. After initialization, a time equal to a period of the driving oscillator in the first array will be spent in order to determine if a bit equal to `$0$' or `$1$' passes by site $L _0$. This determination is done by comparing the local-energy amplitude of site $L _0$ with respect to a predetermined cutoff limit. Next, a driving amplitude in the first array defined by (\ref{Eqn:AmplitudeMult}) will be used for a period of time equal to $P _1$, and the process continues inductively.

\begin{figure*}
\centerline{
\begin{tabular}{cc}
{\small \bf Array $0$} & {\small \bf Array $1$} \\
\includegraphics[width=0.47\textwidth]{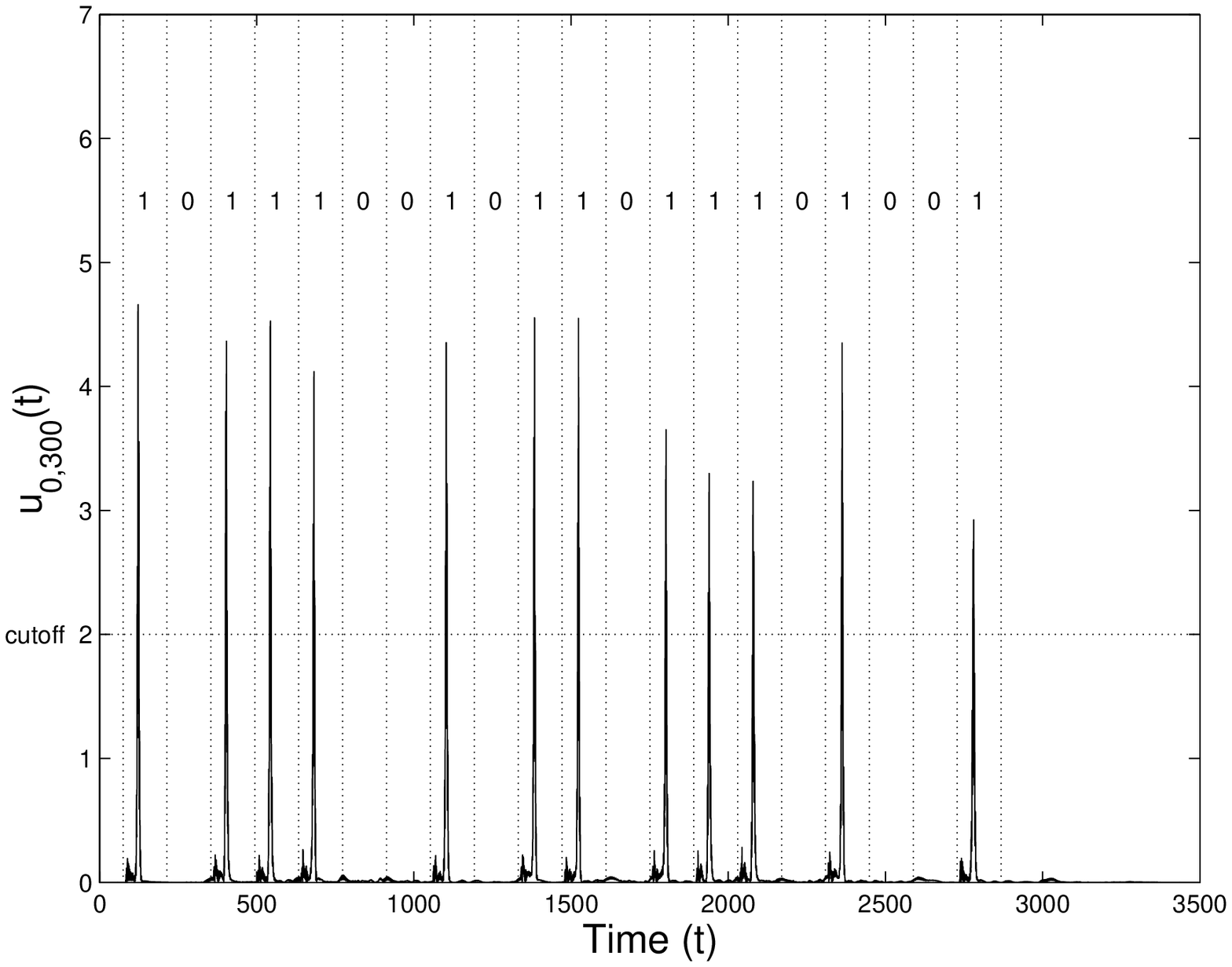}&
\includegraphics[width=0.47\textwidth]{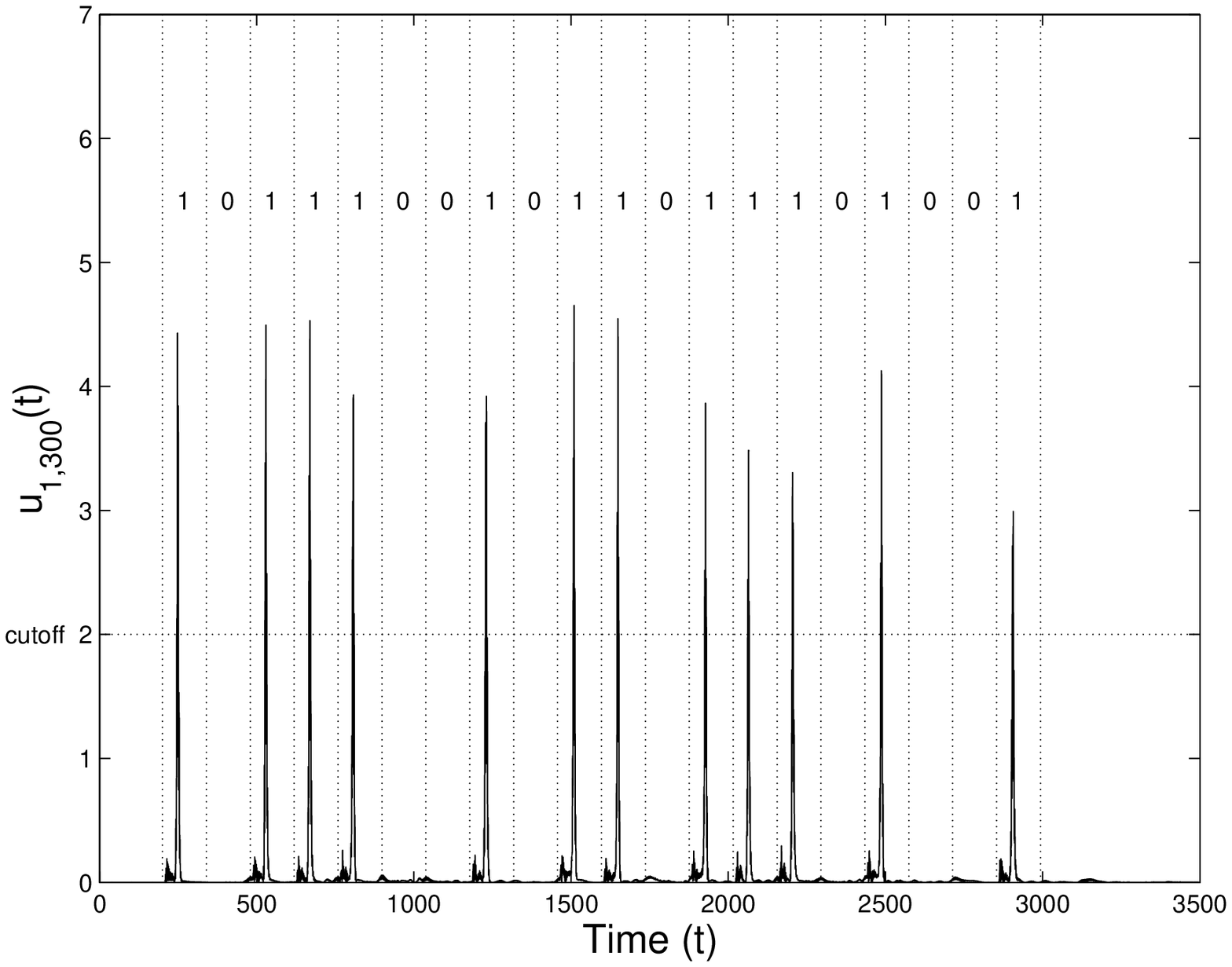}\\
{\small \bf Array $2$} & {\small \bf Array $3$} \\
\includegraphics[width=0.47\textwidth]{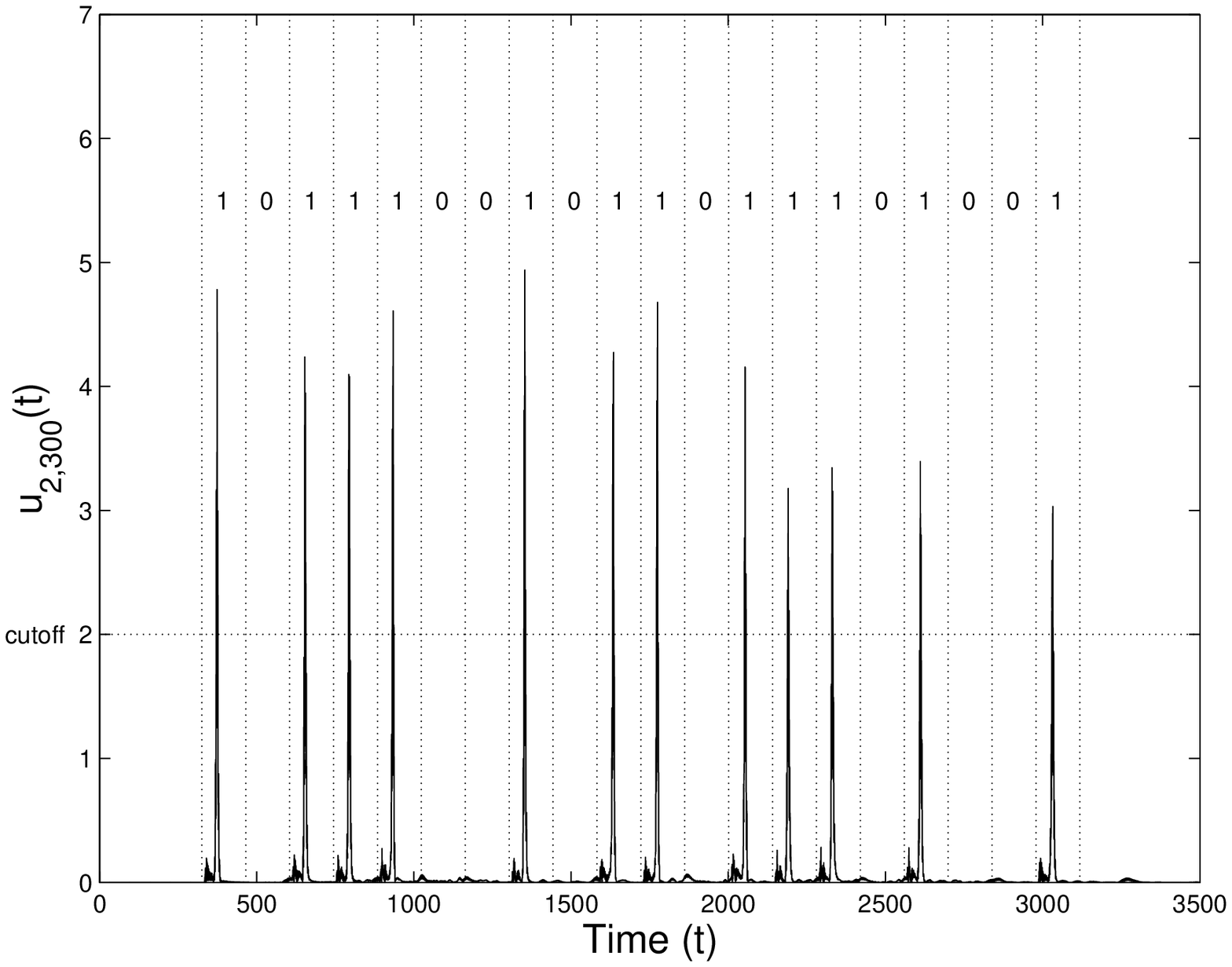}&
\includegraphics[width=0.47\textwidth]{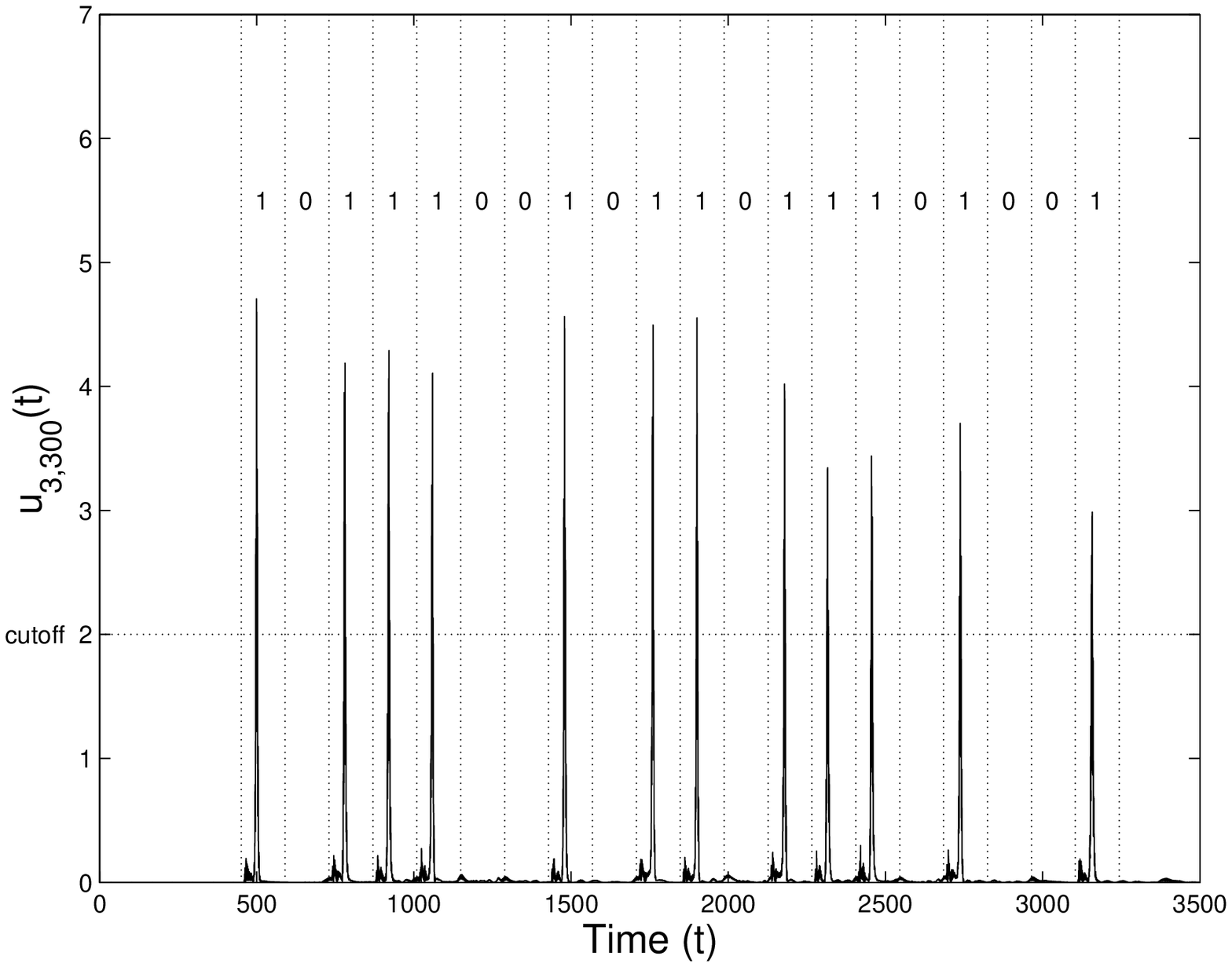}\\
{\small \bf Array $4$} & {\small \bf Array $5$} \\
\includegraphics[width=0.47\textwidth]{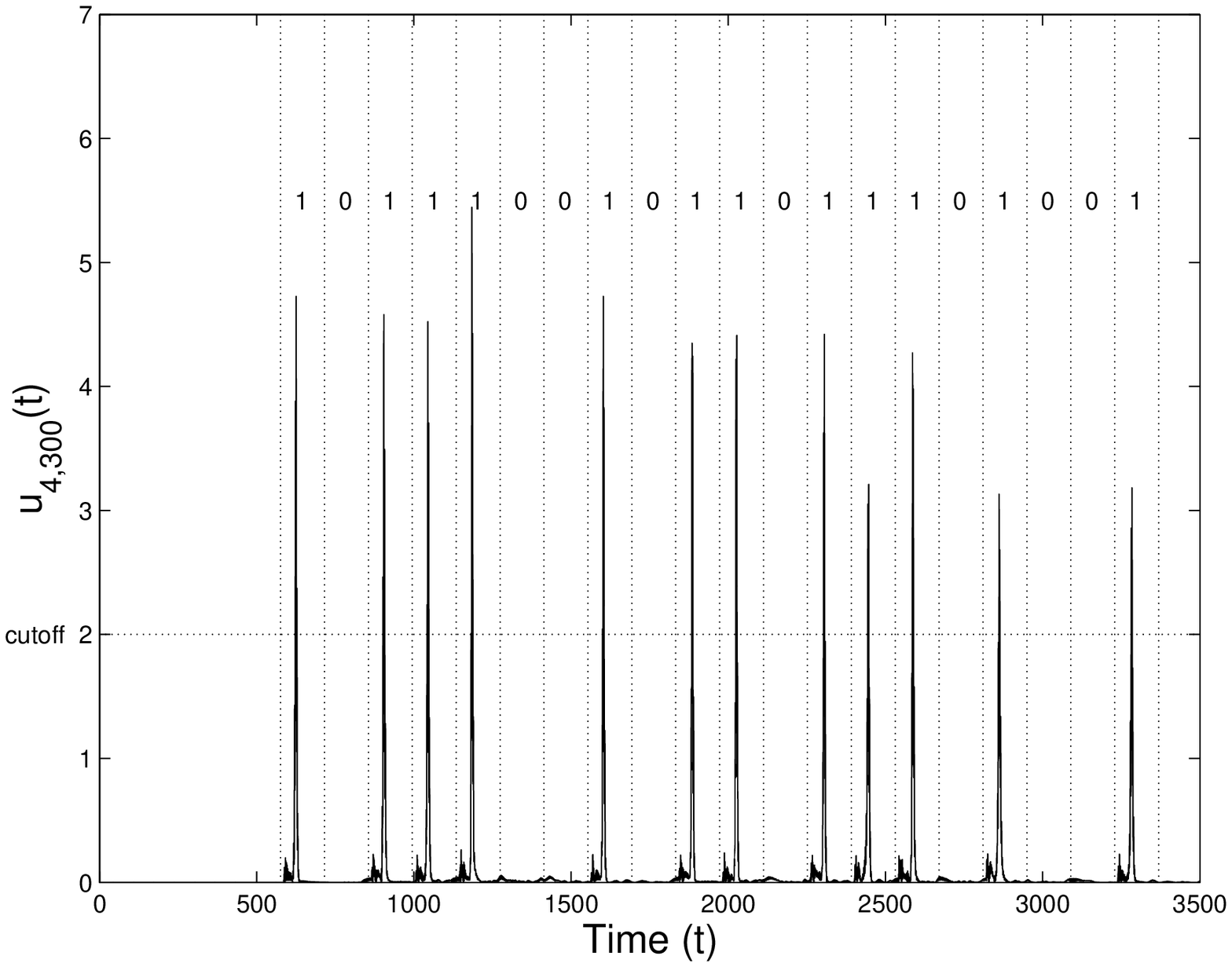}&
\includegraphics[width=0.47\textwidth]{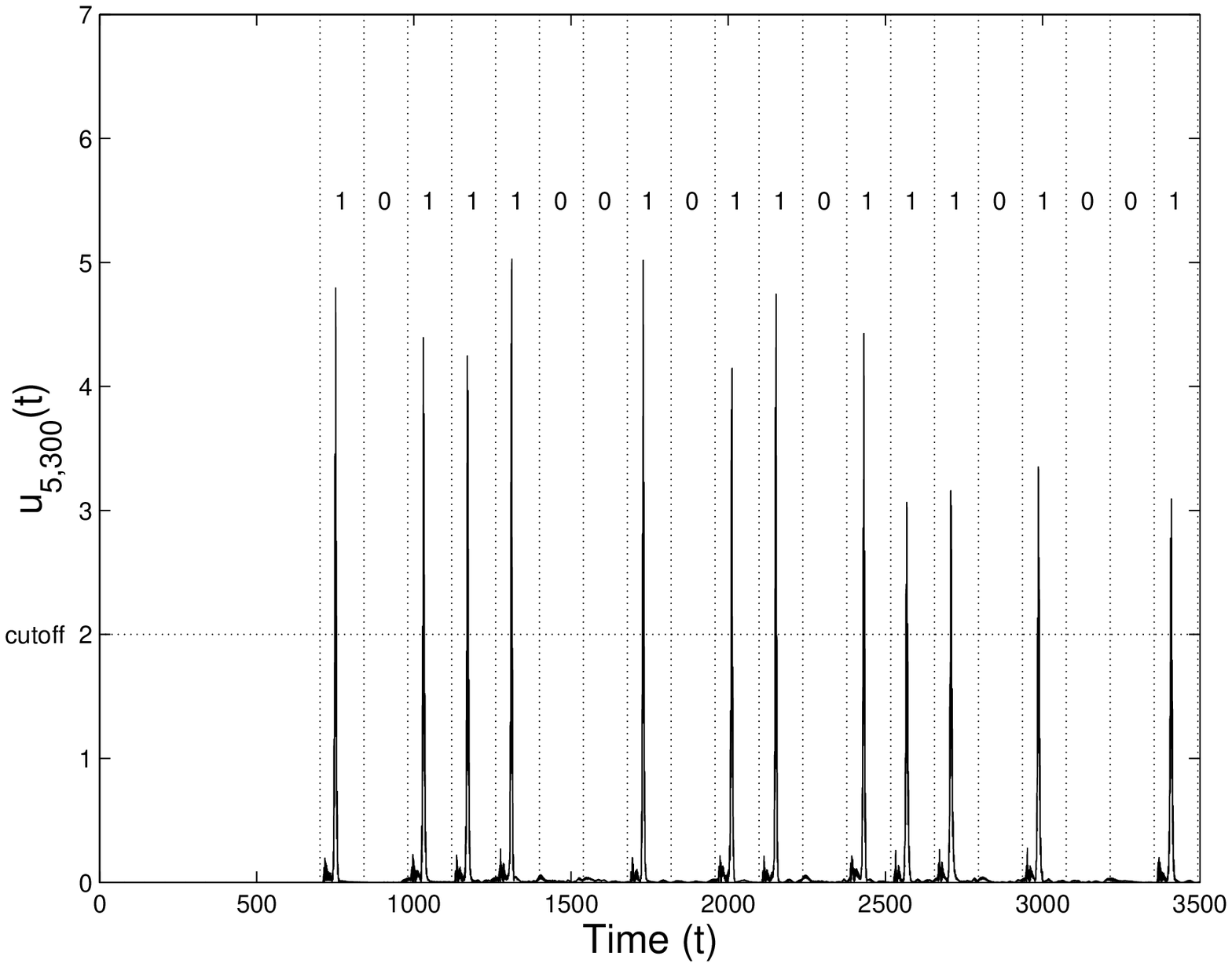}
\end{tabular}
}
\caption{Time-dependent local energies at sites $300$ in a linear concatenation of mechanical arrays of length $5$, as a result of the transmission of signal `$10111001011011101001$' and using a cutoff limit equal to $2$. The decoding of the transmitted message is shown in each array at the top of the reception periods. \label{Fig:Simulation}}
\end{figure*}

Following the conventions of Section \ref {NumScheme}, the $j$-th system of differential equations in model (\ref {Eqn:ApplicationEq}) will be approximated numerically by a finite number of equations $N _j \gg L _j$ via the set of finite-difference schemes 
\begin{equation}
\displaystyle {\frac {\delta ^2 _t u _{j n} ^k} {(\Delta t) ^2} - \left( c _j ^2 + \frac {\alpha _j} {2 \Delta t} \delta _t \right) \delta ^2 _x u _{j n} ^k + \frac {\beta _j ^\prime} {2 \Delta t} \delta _t u _{j n} ^k + \frac {V (u _{j n} ^{k + 1}) - V (u _{j n} ^{k - 1})} {u _{j n} ^{k + 1} - u _{j n} ^{k - 1}}} = 0,
\end{equation}
where $\beta _j ^\prime$ is the sum of the external damping coefficient in the $j$-th mechanical array and the effect of an absorbing boundary in the last few lattice sites. We mimic now the rest of Section \ref {NumScheme} in order to define computational schemes to approximate the total energy and the individual energies in each mechanical array.

\subsection{Simulation}

As an application, assume that the binary signal `$101110010110\-11101001$' will be transmitted in a linearly concatenated, mechanical system (\ref {Eqn:ApplicationEq}) of length $5$, in which a period of single-bit generation equal to $20$ driving periods will be used. For the sake of simplicity, we will assume that $\alpha _j = 0.02$, $\beta _j = 0.003$, $\gamma _j = 0.01$ and $\Omega _j = 0.9$ for every $j = 0 , 1, \dots , 5$, for which numerical simulations yield a common critical amplitude $A _s = 1.778$ and a phase velocity equal to $2.395$.

Computationally, we consider arrays consisting of $N = 600$ sites each, and assume that the link between two consecutive arrays is invariably placed on site $L = 300$. An analysis similar to that done in Section \ref{Sec3} reflects that the estimate for the minimal value of the energy carried by a breather at site $L$ is above the value of $A _s$ when an amplification coefficient $C = A _s$ is used. 

In this state of matters, Figure \ref {Fig:Simulation} presents the time-dependent local energy of the $300$-th site in each of the $6$ mechanical arrays in our system, as a consequence of the transmission of signal `$10111001011011101001$'. For the sake of convenience, each graph shows approximate periods of reception and, the value of the transmitted bit on top of each period. Our numerical results establish that a rather reliable transmission can be achieved using model (\ref {Eqn:ApplicationEq}) under suitable parametric conditions.

\section{Conclusions and perspectives\label{Sec5}}

In this work, we have presented a simple, mathematical model to efficiently transmit binary signals in certain semi-infinite, discrete mechanical systems using the process of nonlinear supratransmission. The model is physically described as a concatenation of a finite sequence of $n$ such arrays, in such a way that the amplitude of the driving oscillator at the end of the $k$-th lattice for $2 \leq k \leq n$, is equal to the local energy amplitude of some fixed oscillator in the $(k - 1)$-st lattice, located relatively far away from its driving oscillator. Of course, this energy-bound series of arrays requires that the local energy of the linking site in the $(k - 1)$-st lattice induced by the moving breather be at least equal to the critical amplitude of the driving disturbance in the $k$-th lattice, and simulations of this type of structures have shown that perfect transmission may be achieved under suitable choices of parameters.

As a means in order to procure a better transmission, this work presents bifurcation diagrams of occurrence of nonlinear supratransmission in non-ideal situations. Thus, numerical predictions of this process have been provided in the presence of external and internal damping, and nonzero values of what we have called normalized bias currents. In the former case, a delay in the appearance of the critical amplitude at which supratransmission starts is observed, as well as a decrease in the region where harmonic phonon quenching takes place. In the case of external damping, an expansion in the forbidden band gap region and a delay on the occurrence of the supratransmission threshold are clearly noticed. On the other hand, the normalized bias current tends to rush the appearance of supratransmission. Likewise, our results show that the localized structure produced in our model move away from the boundary at constant speeds, even in the presence of external or internal damping, or nonzero normalized bias current. 

Before closing this paper, it is important to recall that our mathematical model (which is, after all, a Dirichlet boundary-value problem with zero initial conditions) propagates localized energy in the form of moving breathers. Naturally, question arises as to how different the propagation of energy will be in the case of a Neumann boundary-value problem, where one immediately recognizes a system of Josephson junctions coupled through superconducting wires. In this case, the existence of a nonlinear supratransmission threshold has been analytically and numerically established for an undamped system without normalized bias current \cite{Chevrieux2}. Moreover, the existence of a bifurcation value under which energy transmission is drastically suspended---a process we have called \emph {nonlinear infratransmission} in an upcoming paper---has been reported in the same work; however, no bifurcation analysis has been given for it yet, not even in the ideal case.

In view of these facts, it is particularly interesting to describe numerically the occurrence of the nonlinear supratransmission and infratransmission thresholds under the presence of damping and nonzero normalized bias currents. Moreover, it is indispensable to determine the nature of the localized structures transmitted into discrete Josephson-junction arrays by the driving boundary. Preliminary results have shown that energy is transmitted into these systems in the form of kinks and anti-kinks, the properties of which are still important topics of study.

Another interesting direction of investigation arises from the fact that certain type of nonlinear differential equations (amongst which is the sine-Gordon model \cite{Faber}) admit localized solutions that stabilize by dissipative effects \cite{Kundu}. These types of solutions have been studied in continuous models, thus the existence solutions in a discrete system of coupled sine-Gordon equations would also be an interesting question to attack. 

\subsubsection*{Acknowledgment}
The authors wish to express their gratitude for the comments emitted by the referees. One of us (JEMD) wishes to thank Dr. \'{A}lvarez Rodr\'{\i}guez, dean of the Centro de Ciencias B\'{a}sicas of the Universidad Aut\'{o}noma de Aguascalientes, and Dr. Avelar Gonz\'{a}lez, head of the Direcci\'{o}n General de Investigaci\'{o}n y Posgrado of the same university, for uninterestedly providing him with the means to produce this paper. The present work represents a set of partial results under project PIM07-2 at this university.

\end{document}